\documentclass[fleqn,usenatbib]{mnras}
\usepackage{newtxtext,newtxmath}
\usepackage[T1]{fontenc}
\usepackage{ae,aecompl}
\usepackage{graphicx}	
\usepackage{amsmath}	
\usepackage{amssymb}	
\newcommand{\etaeta} {\eta/\eta_*}

\newcommand{\rph} {r_{\rm ph}}
\newcommand{\rs} {r_{\rm s}}
\newcommand{\be}{\begin{equation}}
\newcommand{\ee}{\end{equation}}
\newcommand{\Ll}{\ensuremath{\widetilde{L}_{\widetilde{\nu}}}}
\newcommand{\Lintl}{\ensuremath{\widetilde{L}}}
\newcommand{\Fl}{\ensuremath{\widetilde{F}_{\widetilde{\nu}}}}
\newcommand{\Il}{\ensuremath{\widetilde{I}_{\widetilde{\nu}}}}
\newcommand{\mul}{\ensuremath{\widetilde{\mu}}}
\newcommand{\D}{\ensuremath{\,\mathrm{d}}}
\newcommand{\Omegal}{\ensuremath{\widetilde{\Omega}}}
\newcommand{\nul}{\ensuremath{\widetilde{\nu}}}
\newcommand{\I}{\ensuremath{I_\nu}}

\usepackage{ulem}

\usepackage{color}

\title[Emission from accelerating jets in GRBs]{Emission from accelerating jets in gamma-ray bursts: Radiation dominated flows with increasing mass outflow rates}
\author[F. Ryde et al.]{Felix Ryde$^{1,2}$\thanks{E-mail: fryde@kth.se},  Christoffer Lundman$^{3}$, Zeynep Acuner$^{1,2}$
\\
$^{1}$Department of Physics, KTH Royal Institute of Technology, AlbaNova, SE-106 91 Stockholm, Sweden\\ 
$^{2}$The Oskar Klein Centre for Cosmoparticle Physics\\
$^{3}$Physics Department and Columbia Astrophysics Laboratory, Columbia University, New York, NY 10027, USA
}
\date{Accepted XXX. Received YYY; in original form ZZZ}
\pubyear{2017}

\hypersetup{draft}
\begin{document}
\label{firstpage}
\pagerange{\pageref{firstpage}--\pageref{lastpage}}
\maketitle

\begin{abstract}
We study the narrowest spectra expected from  gamma-ray bursts.
We present an analytical function for the spectrum that is emitted from the photosphere of a radiation-dominated flow that is under acceleration. This is the narrowest possible spectrum and it differs from a Planck function. We also present numerical spectra from photospheres occurring during the transition into the coasting phase of the flow. Using these spectral models, we reanalyse Fermi observations of GRB100507 and GRB101219, which both have been reported to have very narrow spectra.
The bursts can be fitted by the spectral models: For GRB101219 the spectrum is consistent with the  photosphere occurring below or close to the saturation radius, while for GRB100507 the photosphere position relative to the saturation radius can be determined as a function of time. In the latter case, we find that the photosphere initially occurs in the acceleration phase and thereafter transitions into the coasting phase.
We also find that this transition occurs at the same time as the change in observed cooling behaviour: the temperature is close to constant before the break and decays after. 
We argue that such a transition can be explained by an increasing mass outflow rate. 
Both analysed bursts thus give strong evidence that the jets are {(initially)} radiation dominated.
\end{abstract}

\section{Introduction}

Many gamma-ray burst (GRB) spectra 
{have narrow $\nu F_\nu$ peaks} compared to what is expected from non-thermal emission \citep{Axelsson2015,Yu2015}. {Furthermore, a large fraction of GRBs have a low-energy photon index which is harder than allowed by optically-thin emission models \citep{Preece1998}.} Therefore, attention has been drawn to the photosphere of the GRB jet \citep[e.g.][]{Meszaros&Rees2000, Rees&Meszaros2005, Peer2007, Thompson2007} in order to explain the observed spectral shapes and evolutions \citep[e.g.][]{Ryde2004, RydePeer2009, Guiriec2011}. At the photosphere the jet becomes transparent and photons stream freely to the observer. {The observed spectrum is a superposition of Doppler boosted local comoving frame spectra emitted from the photosphere.}

Indeed, GRB jets are thought to be highly opaque close to the central engine, and therefore the photon spectrum at the base of the jet is expected to have
the shape of a blackbody. In a radially expanding, passively cooling jet, that is, a jet with no shocks or other forms of energy dissipation, the
relic thermal radiation field is simply advected out with the flow, while keeping coupled to the electrons. However, even though the initial comoving spectrum is blackbody, the observed spectrum may have a broader shape. This is due to several reasons. First, the angular distribution of the photons in the lab frame will be affected by the radial expansion, and they typically become more anisotropic the closer the photosphere they reach \citep{Beloborodov2010}. Second, photons make their last scatterings at different radii, covering more than an order of magnitude in range \citep{Peer2008,Beloborodov2010,begue2013monte,Lundman2013}. If the observed temperature of the plasma is varying with radius (e.g. though adiabatically cooling) the observer would  detect a multi-colour blackbody (e.g. used for analysing bursts in \citet{Ryde2010}, Mikato et al. 2017). Third, the photons also make their last scatterings at different angles to the observer line-of-sight and are thus Doppler boosted with different strengths \citep{Abramowicz1991,Peer2008,Lundman2013} and, likewise, the observed spectrum will be broadened. In addition to these effects,  if significant energy dissipation occurs in the jet, the thermal equilibrium in the comoving frame can be destroyed and many types of (broad) spectra would be observed (\citet{Rees&Meszaros2005}, \citet{Peer&Waxman2005}; \citet{Peer2006}; \citet{Giannios2006, Giannios2008}; Ioka et al. 2007; \citet{Beloborodov2010}; \citet{Lazzati2011} and used for analysing bursts  in e.g. \citet{ahlgren2015confronting}, Vianello et al. 2017). Finally, spectral analysis requires finite time-intervals to be used,  which could lead to noticable broadening, if there is significant spectral evolution within the analysed time-interval \citep[e.g.][]{Ryde1999, Burgess2014}. Broadening of the observed spectrum, by any of these effects, is thus highly likely.  Therefore, detecting very narrow spectra yield strong constraints on the processes leading to their emission.

GRB jets can be accelerated either by radiation pressure, if the magnetic fields are subdominant (thermal acceleration) or, in the case of Poynting flux dominated flows, by, e.g., magnetic reconnection (magnetic acceleration). In the latter case, continuous dissipation occurs of the magnetic energy, which is converted into kinetic energy of the bulk outflow motion. The acceleration then occurs at a slower rate ($\Gamma \propto r^{1/3}$, where $r$ is the radius from the central engine) than in the radiation pressure case ($\Gamma \propto r$). The photospheric emission in Poynting flux dominated models are expected to have broad spectral shapes \citep{Giannios2006} and to have high peak energies due to photon starvations \citep{BeguePeer2015}. 

The narrowest spectra should thus be found in thermally accelerated jets where the photosphere occurs during the acceleration phase, before it saturates at its final Lorentz factor, $\Gamma$. This is because the flow is then radiation dominated (photon energy density dominates over kinetic energy density) and, in addition,  energy dissipation, such as internal shocks are not expected. Therefore, the photon spectral distribution can be expected to be maintained.  Moreover, in the case of thermally accelerated flows ($\Gamma \propto r$) the temperature in the  observer frame is constant \citep{Meszaros&Rees2000}. Therefore, any distribution in radii and angles of the last scattering sites would not contribute to a broadening of the spectrum. Finally, to ensure a narrow spectrum,  the flow can only be weakly dissipative, preventing significant amount of energy being injected in the flow changing the electron distribution and dynamics of the flow. 

In particular, \citet{Beloborodov2011} showed that, during the acceleration phase, the photospheric spectrum (in the local comoving frame) is exactly a Planck function, if it initially was one. The reason for this is that two effects compensate each other: First, the angular distribution of the radiation field, in the lab frame, changes between scattering  points due to the spherical expansion \citep{Beloborodov2010}. 
Second, in the case of  a flow with $\Gamma \propto r$ the Lorentz transformation of angles into the local comoving frame will exactly compensate this change. Therefore, the photon angular distribution and spectral shape  in the local comoving frame stays intact.

In this paper, we study the shape of the narrowest spectra since these are undisputably from the photosphere. First, we investigate the spectra shapes that are expected from photospheres occurring at different radii relative to the saturation radius (\S \ref{sec:spectra}). After deriving the spectral shapes, we fit the spectra of two bursts observed by the {\it Fermi Gamma-Ray Space Telescope} (\S \ref{sec:fits}). Finally, we discuss consequences of our analysis for the cooling and structure of the emitting plasma \S\ref{sec:sss}. 

\section{Spectrum from the GRB photosphere}
\label{sec:spectra}

In the fireball model of GRBs the initial dynamics of the jet is determined by its thermal pressure.
At the high temperatures in the jet, photons outnumber { other particles (electrons/positrons/protons) by a large factor and therefore the pressure is dominated by the photon pressure. As a consequence,} the jet Lorentz factor initially increases at the expense of the jet internal energy.  The Lorentz factor may saturate either due to the fact that photons escape (the photosphere occurs during the acceleration phase), or that the internal energy becomes too low to accelerate the outflow further. The latter occurs at the saturation radius ($r_{\rm s}$), essentially dividing the jet flow into an acceleration phase ($r << r_{\rm s}$) and a coasting phase ($r >> r_{\rm s}$)\footnote{We note that the transition is gradual, but nontheless $r_{\rm s}$ is a useful parameter}.

\subsection{Transparency in the coasting phase}

Typically, the photosphere is assumed to occur in the coasting phase. The reason is that in many bursts a strong non-thermal component is observed in addition to the photospheric component \citep[e.g.][]{Ryde2005, RydePeer2009, Guiriec2010, Guiriec2011, Ryde2010, Axelsson2012, Burgess2011, Iyyani2016}. Since the kinetic energy density dominates over the photon energy density in this phase, there is enough energy to be dissipated at a larger radius ($r>>r_{\rm ph}$) to produce a significant and observable, optically-thin, non-thermal emission component. This component is often interpreted as synchrotron emission \citep[e.g.][]{Meszaros&Rees2000, Ryde2005,Battelino2007}. 

\begin{figure}
	\includegraphics[width=\columnwidth]{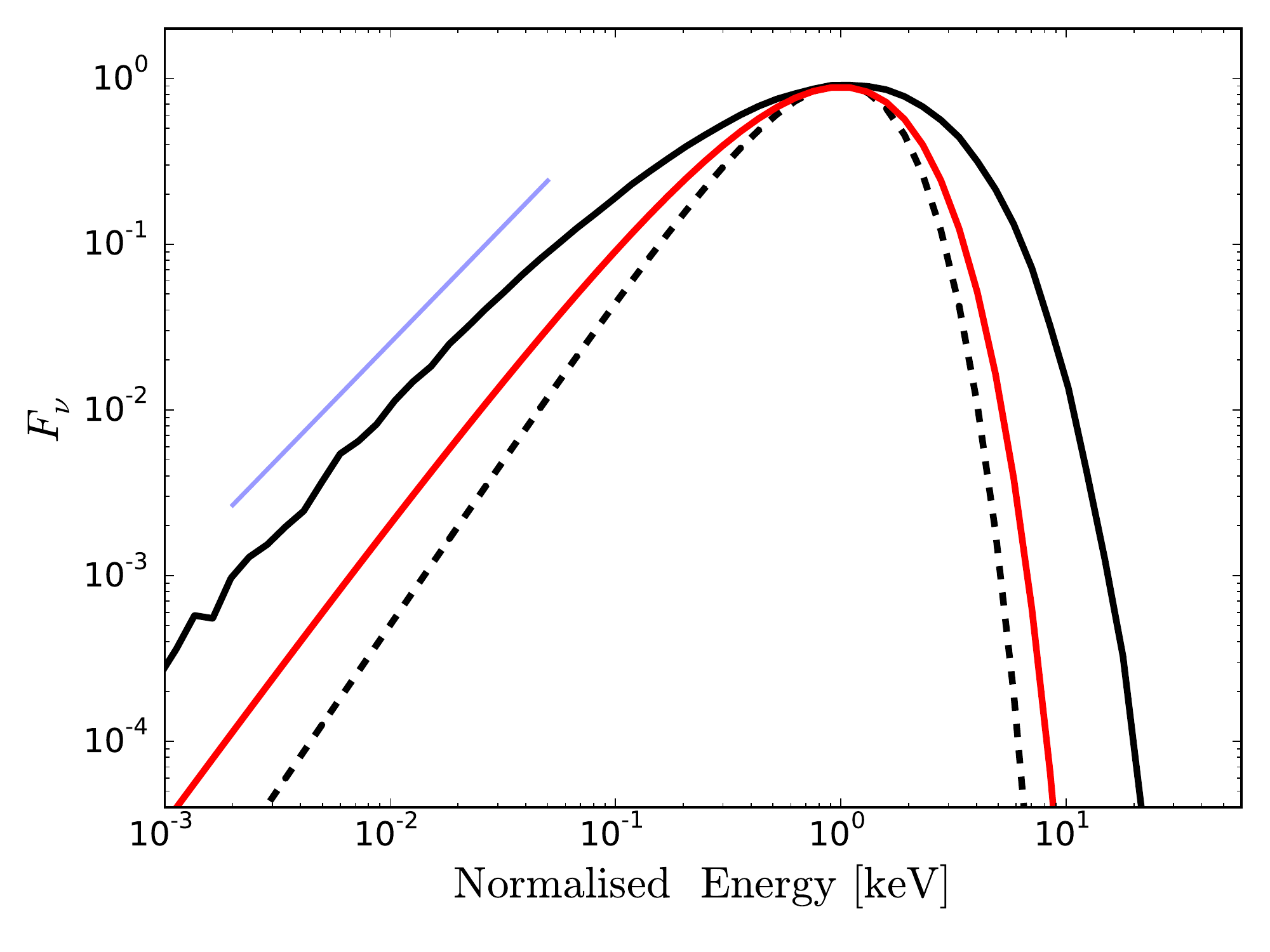}
    \caption{Spectra expected from the photosphere in a non-dissipative 
GRB jet.  The black, solid line is the spectrum from the photosphere occurring during the coasting phase in a passively cooled jet. The blue line shows a power-law with a photon-index $\alpha = 0.4$. The red, solid line is the spectrum from the photosphere occurring during the acceleration phase (\ref{eq:aBB}). The blackbody (eq. \ref{eq:BB}) is shown as a reference by the black, dashed line.   Note that a pure blackbody is never expected from a GRB.}
    \label{fig:1}
\end{figure}

The narrowest possible spectrum in the coasting phase, is from a passively cooled jet without any energy dissipation that can alter the spectrum. This spectral shape was first derived by \citet{Beloborodov2010} \citep[see also][]{Lundman2013, Ito2013, begue2013monte}. The spectrum is significantly broader than a blackbody and the low-energy slope has a photon index of $\alpha \sim 0.4$. The main contribution to the broadening is due to the decrease in Doppler boost for photons which decouple at angles $\gtrsim 1/\Gamma$ with respect to the observer line-of-sight. In Fig. \ref{fig:1} such a spectrum is shown by the black, solid line.

\subsection{Transparency in the acceleration phase}


As mentioned in the introduction the local comoving-frame spectrum will remain blackbody and isotropic throughout the acceleration phase as long as $\Gamma \propto r$. Its spectral shape is given by the Planck function (here given as a normalised $\nu L_{\rm \nu}$ function)
\begin{equation}
    \frac{\nu L_\nu}{L} = \frac{15}{\pi^4}\frac{x^4}{\exp{x}-1},
	\label{eq:BB}
\end{equation}
where $x = h\nu / \Gamma k T$ is the photon energy normalized to the boosted
temperature), $h$ is the Planck's constant and $k$ is the Boltzmann's constant. The Rayleigh-Jeans law ($x << 1$) gives a power-law slope of $x^3$, which corresponds to a photon index of $\alpha = 1$. The spectrum is plotted as the dashed, black line in Figure \ref{fig:1}.

In order to compare to observations, we need to boost the comoving blackbody spectrum into the lab frame. As shown in appendix A an analytical solution can be found in the case where $\Gamma \propto r$ (which is valid far below the saturation radius). The resulting lab frame spectrum is given by 
\begin{equation}
    \frac{\nu L_\nu}{L} = \frac{45}{8\pi^4}\, x^3\,\left( \frac{x}{2} - \ln \left[ \exp{\frac{x}{2}} - 1 \right]
   \right).
	\label{eq:aBB}
\end{equation}
Equation (\ref{eq:aBB}) is the narrowest spectrum that possibly can be observed from a GRB. It is narrower than the corresponding spectrum emitted in the coasting phase, due to the fact that photons which make their last scattering at large angles ($\gtrsim 1/\Gamma$) do not suffer such a large decrease in the Doppler boost. Photons which decouple at larger angles also decouple at larger radii (due to the angular dependence of the optical depth), where the flow has a larger Lorentz factor (since $\Gamma \propto r$) which helps to compensate the decrease in Doppler boost due to the larger angle.

Equation (\ref{eq:aBB}) is plotted as the red, solid line in Figure \ref{fig:1}. The $\nu L_\nu$ peak is at $5.83\, kT$ compared to $3.92\, kT$  for the Planck function\footnote{Corresponding value for the $F_\nu$ peak is $3.66 \, kT$ compared to the Planck function's $2.82 \, kT$}. In the limit of $x << 1$, the spectrum follows $x^3 \log x$, which is slightly shallower than the Rayleigh-Jeans' slope of the blackbody. 

Finally, we note that if a spectrum is detected having the shape of Equation (\ref{eq:aBB}), a strong requirement is put on the radiative efficiency, which must be larger than $\sim 0.5$, since $r_{\rm ph} << r_{\rm s}$.

\subsection{Transition region}

The saturation radius is $r_{\rm s} \sim \eta r_0$, where $\eta$ is the dimensionless entropy of the flow $\eta = L/\dot{M} c^2$ while the photospheric
radius in the coasting phase is given by $r_{\rm ph} \sim  L\sigma_{\rm T}/4 \pi m_{\rm p} c^3 \eta \Gamma^2$.  Since a saturated jet has $\Gamma = \eta$, we find by setting $r_{\rm s} = r_{\rm ph}$ that 
\begin{equation}
    \eta_* =  \left( \frac{L \sigma_{\rm T} }{4 \pi m_{\rm p} c^3 r_0} \right)^{1/4},
	\label{eq:etastar}
\end{equation}
which is the critical value of $\eta$ (for a given $L$ and $r_0$); if $\eta < \eta_*$ then $r_{\rm ph} > r_{\rm s}$ and thus occurs in the coasting phase, while if $\eta > \eta_*$ then $r_{\rm ph} < r_{\rm s}$ and thus occurs in the acceleration phase.  In reality, the end of the acceleration phase occurs over more than
one decade in radius and therefore has a smooth transition between $\Gamma \propto r$ to  $\Gamma = {\rm const}$.  

The exact details of the observed spectral shape in this transition region depends on the Doppler boost at the final scatterings, and thus on how the Lorentz factor scales with radius, which in turn depends on when the photons decouple from the flow.  In order to find the observed spectra, simulations are therefore needed. For this reason we self-consistently simulate the radiative transfer and the  flow-dynamics. We use a code which couples 1D hydrodynamics to Monte Carlo photon propagation (Lundman et al. 2017, in prep.). The photons act as sources of energy and momentum for
the hydrodynamics, which means that they accelerate the flow if they carry
enough energy and the flow is optically-thick, and gradually decouple if the flow becomes optically-thin. The decoupling could then lead to the 
flow saturating at a lower Lorentz factor than $\eta$.

We simulate flows with different ratios of $\etaeta$. For all simulations we
choose $L = 10^{52}$ erg/s and $r_0 = 10^7$ cm, but these specific values are
not important; what matters is $\etaeta$, as this ratio holds information on the location of the photosphere relative to the saturation radius. We consider
a range of $\etaeta$ which ensures that the limits $\rph >> \rs$  and $\rph << \rs$ are included, along with points in between. {The simulations start well below the saturation radius, and follow the flow and radiation until the spectrum stops changing, far above the photosphere.}


\begin{figure}
	\includegraphics[width=\columnwidth]{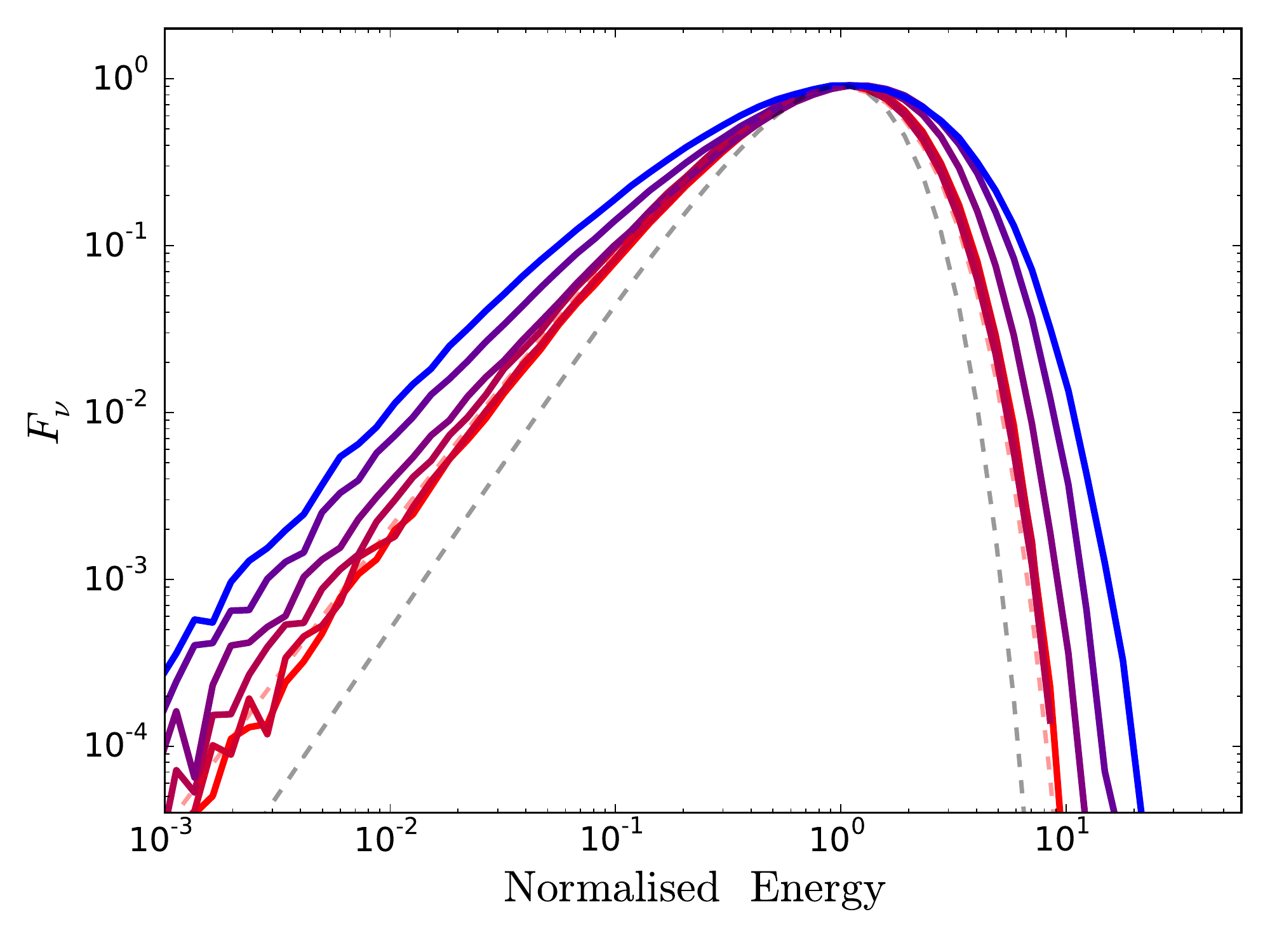}
    \caption{
The colored, solid lines are for spectra with varying values of $\eta/\eta_*$:  $100$ (red), $31.6$, $10$, $3.16$, $1$, $0.316$ (blue). 
Finally, the dashed lines are for equation (\ref{eq:aBB}) [red] and equation (\ref{eq:BB}) [black] and are shown for comparison.}
    \label{fig:trans1}

\end{figure}

Figure \ref{fig:trans1} shows the observed spectra for different positions of the $\rph$ relative to the $\rs$ (different values of $\etaeta$). As expected, the resulting spectra lie between the acceleration phase spectrum (Eq. \ref{eq:aBB}) and the coasting phase spectrum. 
These spectral models can now be compared to observed GRBs.

\section{Fitting narrow spectra observed by {\it Fermi Gamma-ray Space Telescope}}
\label{sec:fits}

We will now reanalyse the data of two bursts observed by the {\it Fermi Gamma Ray Space Telescope}
that have previously been identified to be extremely narrow: GRB100705 \citep{Ghirlanda2013} and GRB101219 \citep{Larsson2015}. In particular, we want to find out if the width of the spectra are able to constrain the parameter $\etaeta$ and thereby identify the position of the photosphere compared to the saturation radius.

\subsection{Fitting procedure}

We use data from the Gamma-ray Burst Monitor (GBM; \citet{Meegan2009}) on board the {\it Fermi Gamma-ray Space Telescope}. The monitor  consists of 12 sodium iodide (NaI) detectors, that are sensitive between 8 keV and 1 MeV and two bismuth germanate (BGO) detectors, that are sensitive between 150 keV and 40 MeV. We use the time-tagged event data and the standard detector response files, which are provided by the GBM team and are available at the {\it Fermi} Science Support Center\footnote{http:$//$heasarc.gsfc.nasa.gov$/$W3Browse$/$fermi/fermigtrig.html}. We use the most illuminated detectors. The background is estimated from time-intervals before and after the burst and the rate is fitted with polynomial function in time. For the spectral analysis we use the fitting routing 3ML\footnote{http:$//$github.com/giacomov/3ML} \citep{3ML}. We include the analytical function in equation (\ref{eq:aBB}) into the fitting routine and we implement the different numerical spectral shapes for different values of $\etaeta$ into a table model for fitting purposes.

\subsection{GRB100507}
\label{sec:GRB100507}

\begin{figure}
    \includegraphics[width=\columnwidth]{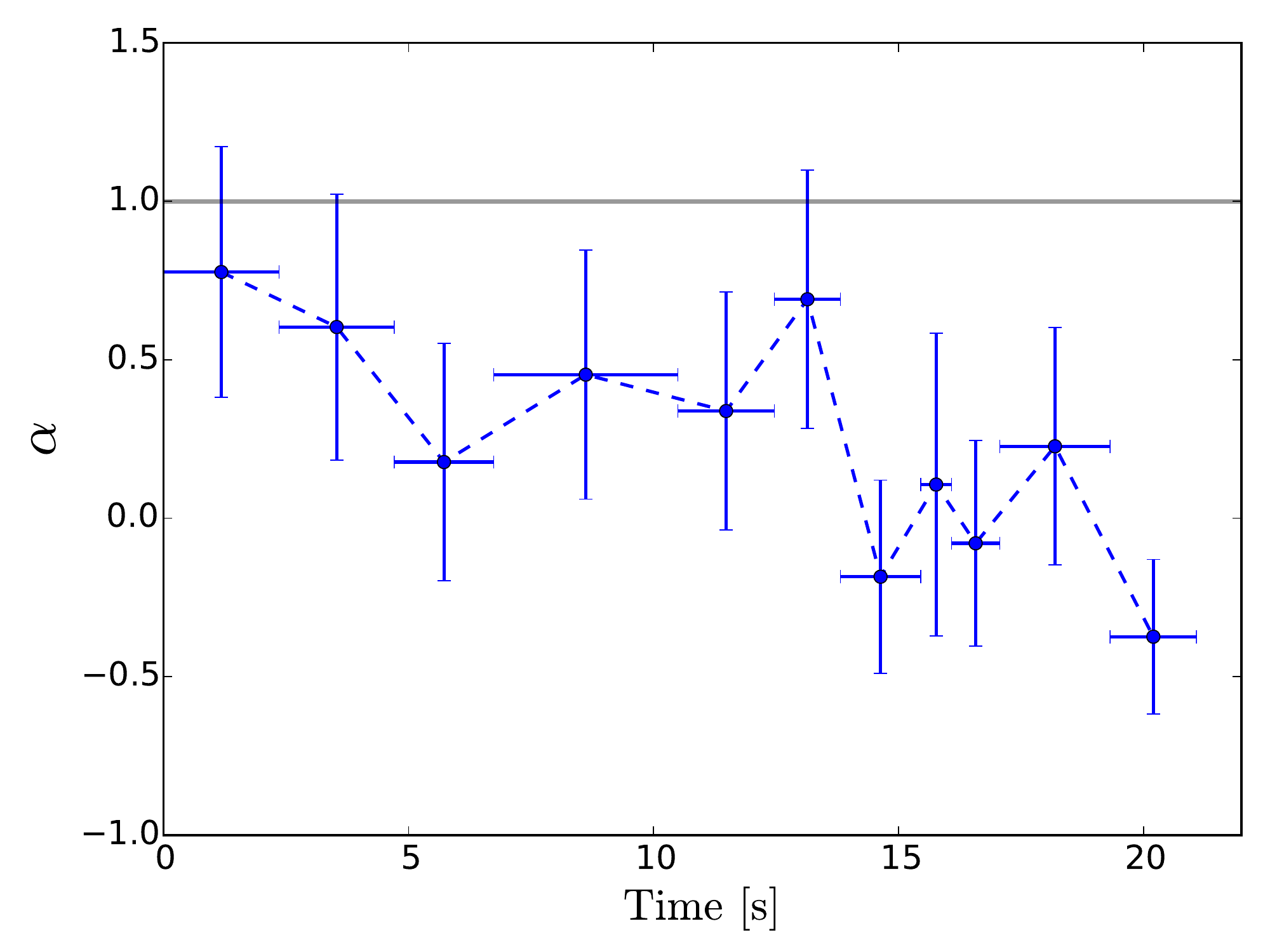}\hfill
	\includegraphics[width=\columnwidth]{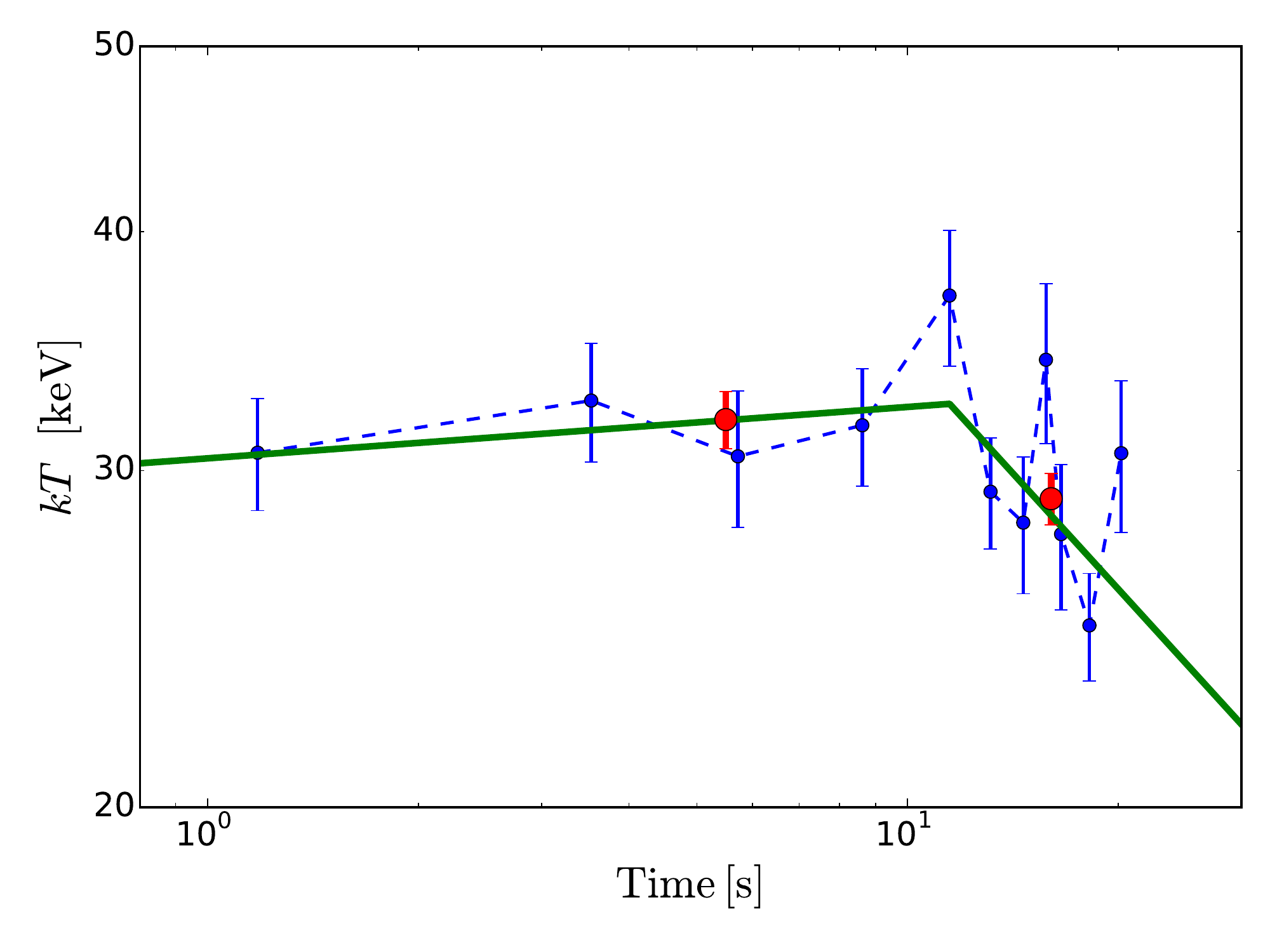}\hfill
	\includegraphics[width=\columnwidth]{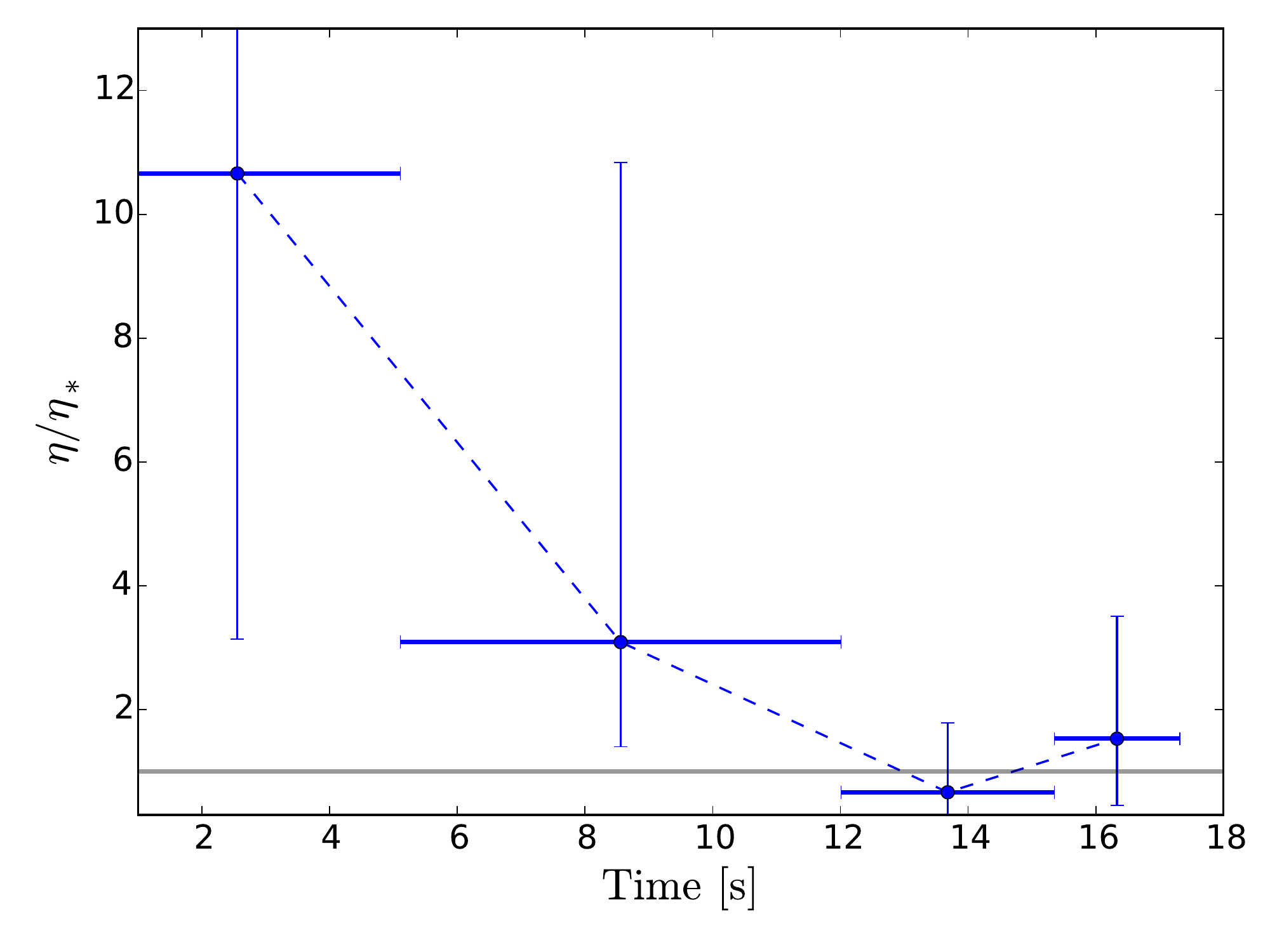}\hfill
    \caption{{\it Fermi}/GBM observations of GRB100507. Upper panel: Low-energy photon index of the Band function (using SNR=10). The grey line corresponds to the Rayleigh-Jeans' slope of a blackbody ($\alpha = 1$). Middle panel: Temperature evolution of blackbody fits (using SNR=10, and a log-log plot). The green line is the best fit for a broken power-law function. Lower panel: Evolution of $\etaeta$ (using SNR=15). The grey line indicates that the spectrum is released at the saturation radius, where $\eta = \eta*$.}
    \label{fig:GRB100705}
\end{figure}

We analyse GRB100507577 over the time range 0 - 30 s relative to the GBM trigger, and use the detectors NaI9, NaI10, and BGO1. 
We first analyse the time-resolved spectra found from requiring a signal-to-noise ratio, SNR = 10 in every time bin. This gives us eleven time bins over the duration 0--21.1 s, adequate to follow the spectral evolution. We fit the standard Band et al. (1993) function, which has four parameters: Two spectral slopes\footnote{Photon index, defined as $N_{\rm E} \propto E^{\alpha}$, where $N_{\rm E}$ is the  photon flux and $E$ is the photon energy.}, $\alpha$ and $\beta$, the peak energy, $E_{\rm p}$, and a normalisation. We find that the low-energy power-law slope $\alpha$, is indeed very hard: $\alpha$ starts off at around 0.7 and decreases to $\alpha \sim -0.25$ as depicted in the upper-most panel in Figure \ref{fig:GRB100705}. As a comparison the  Rayleigh Jeans' slope of the blackbody is $\alpha = 1$ (depicted by the gray line in Fig. \ref{fig:GRB100705}). 

We then fit these spectra with a blackbody (eq. \ref{eq:BB}), which only has two parameters, the temperature, $kT$ and the normalisation. The goodness-of-fit of the blackbody is slightly better than the Band function fits. The temperature evolution, of the blackbody fits, is shown in the middle panel in Figure \ref{fig:GRB100705}. The temperature stays relatively stable and lies between 20 and 40 keV. {Since the cooling in other thermal bursts has been observed to follow a broken power-law decay \citep[e.g.][]{Ryde2004,RydePeer2009, Axelsson2012, Burgess2014a}  and since there is an indication of a break at around 11 s, we attempt at fitting the temperature with a broken power-law function.}
The green line is the resulting best fit (to the 11 data points), with a break time at $11.5 \pm 3$ s, pre-break power-law index $0.03 \pm 0.06$, post-break power-law index $-0.40 \pm 0.023$, and  $\chi^2/{\rm dof} = 11.6/7$. A corresponding fit to a single power law gives $\chi^2/{\rm dof} = 18.1/9$. {Since the $\chi^2/{\rm dof}$ is indeed lower,} we describe the observed temperature evolution as being best described by a broken power-law. Furthermore, the red dots are fits of the temperature for the integrated periods before and after 11 s, reasserting that there is an actual change in temperature. 
 
\begin{figure}
	\includegraphics[width=\columnwidth]{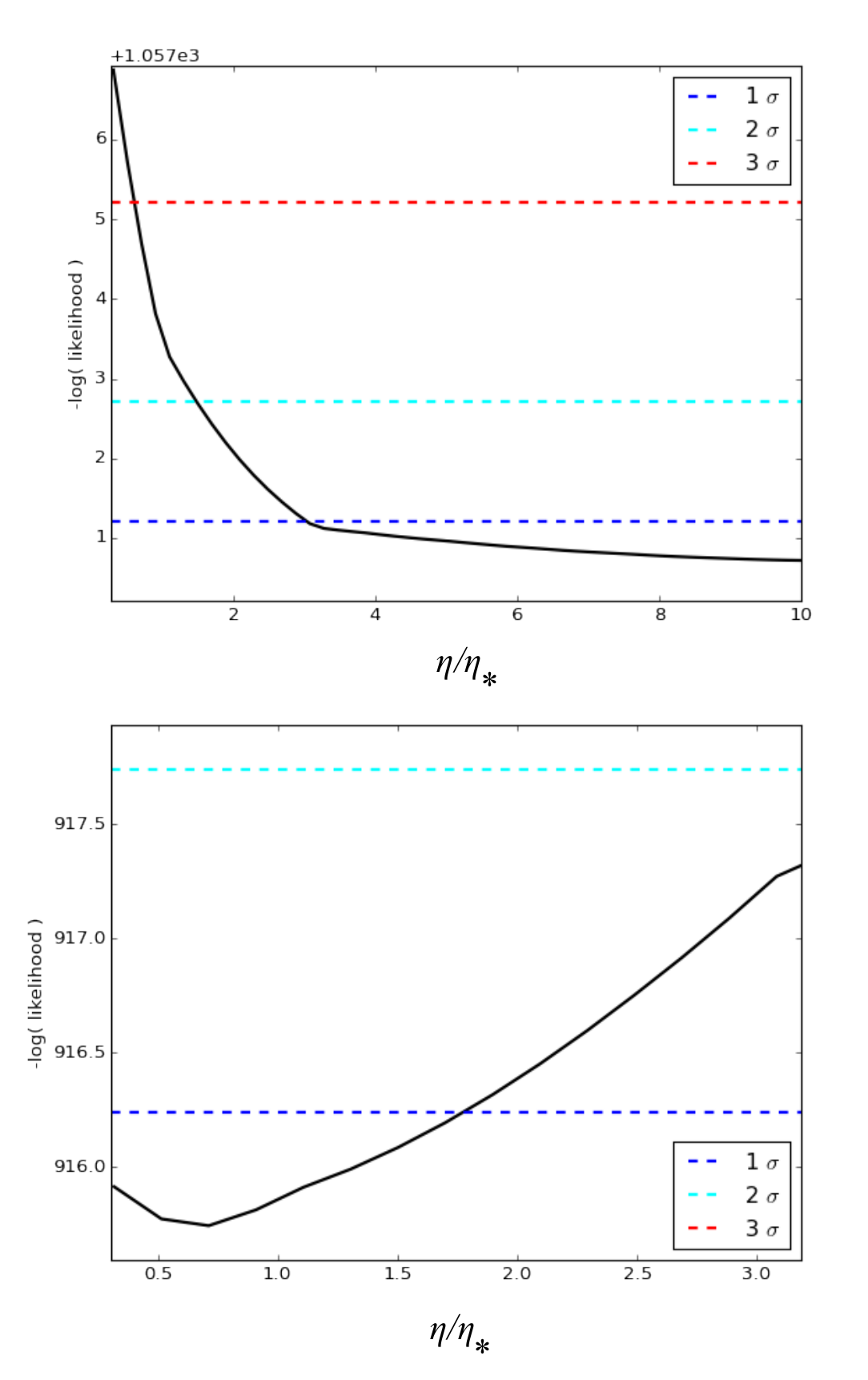}
    \caption{Likelihood profiles of the first and third time bin fits from figure \ref{fig:GRB100705}. The dashed lines indicate the confident levels.}
    \label{fig:lh}
\end{figure}

We now fit the table model that we produced for spectra having different values of $\etaeta$. In addition to $\etaeta$, the fitted parameters are the normalisation and the peak energy. In order to constrain the fits we increase the SNR to 15, which gives us four time-bins (0 -- 17.3 s). 
For these bins the spectra are well fitted. The peak values remain similar to the blackbody fits, while the $\etaeta$ decreases from a large value with {a lower limit of}  $\etaeta \sim 3$ (which makes the spectra similar to that of Equation \ref{eq:aBB}) to values close to and lower than 1. This is shown in the lower-most panel in Figure \ref{fig:GRB100705}. {To illustrate the fits, we show in Figure \ref{fig:lh} the likelihood profiles of the fits to the first and the third time bins. In the first case, the narrowness of the spectrum gives a lower limit on the parameter $\etaeta$, while in the other case it spectrum has broadened such that the fit gives an upper limit of $\etaeta$.}

These fits thus suggest that the narrowness of the spectra and their spectral evolution in GRB100705 (e.g. captured by the Band function) is due to the early spectra being produced below the saturation radius. Eventually, the photosphere approaches the saturation radius. This burst therefore maps out the change in the spectral shape around the saturation radius, leading to the observed broadening. 

We note that the broadening of the late time spectra cannot be explained by subphotospheric dissipation. The reason is that such broadening can only occur during the coasting phase and the broadest spectra that was found in GRB100705 are consistent with the narrowest coasting phase spectra (the black spectrum in Fig. \ref{fig:1}). Therefore, the change in spectral shape must be due to change in $\etaeta$.

During the acceleration phase $L_0$ and $T_0$ of the thermal component are given by the observed values \citep{Meszaros&Rees2000}. Thus we can calculate $r_0$, assuming that the emission is blackbody at the central engine. 
\be
r_0 = \left( \frac{L_0}{4 \pi c a  T_0^4}\right)^{1/2} = \frac{d_{\rm L}}{(1+z)^2}\,\left( \frac{F^{\rm obs}}{ca T^4_{\rm obs} }\right)^{1/2}
\label{eq:ro}
 \ee
In the last step, we used the fact that $L_0= 4\pi d_{\rm L}^2  F^{\rm obs}$ and $T_0 = T^{\rm obs} \, (1+z)$, where $d_{\rm L}$ is the luminosity distance and $z$ is the redshift of the burst. During the acceleration phase in GRB100705 we find that $r_0$ is typically $3 \times 10^9$ cm,  assuming $z=1$. Finally, we point out that the narrowness of the spectrum actually requires Equation (\ref{eq:ro}) to be valid, since the only dynamics that can produce the narrow spectrum is a thermally accelerated flow.


\subsection{GRB101219}
\label{sec:GRB101219}

\citet{Larsson2015} reported a blackbody spectrum throughout the evolution of GRB101219686. The signal in this burst is weak and only two time bins can be analysed (-5--18 s and 18--50 s). Therefore, the detailed temporal evolution cannot  be assessed. We nonetheless reanalyse this bursts with the same time-bins\footnote{We use the detectors NaI3, NaI8, and BGO0}.
In both these bins the difference in likelihood between a blackbody and the spectra presented above is small, and a statistically significant determination cannot be made: only lower limits on the parameter $\etaeta$  can be determined. The 1 $\sigma$ lower limit of the $\etaeta$ parameter is $\sim 0.9$  and $\sim 0.5$ in the first and second time time, respectively. While largely undetermined, the $\etaeta$ values
are consistent with the photosphere occurring below or close to the saturation radius and not necessarily in the pure coasting regime. This is consistent with the finding of \cite{Larsson2015}. Using Equation (\ref{eq:ro}) with the fitted values of $F^{\rm obs}$ and $T_{\rm obs}$, and the value of $z=0.55$, we again find $r_0 = 5 \times 10^9$ cm\footnote{Note that \citet{Larsson2015} assumed a model which includes adiabatic cooling beyond the saturation radius in order estimation of the fireball parameters.}. 

The afterglow measurements of GRB101219 can be used to infer the kinetic energy of the blast wave: $ E_{\rm k} = 6.4 \pm 3.5\times 10^{52}$ ergs. This estimate leads to an observed radiative efficiency of less than 5 \% \citep{Larsson2015}. On the other hand, the narrowness of the spectrum, according to the above analysis, requires the radiative efficiency to be high $\sim 50\%$, which is an apparent contradiction. This point will be discussed in \S \ref{sec:radeff}.

\section{Discussion}
\label{sec:sss}


In the case of GRB100705 there were enough time-bins to study  the spectral evolution, which we found was mainly due to an increase of the ratio $r_{\rm ph}/r_{\rm s}$. Below we argue that this increase is due to increasing mass outflow rate.

We imagine that the central engine continuously supplies the jet with blackbody radiation and that the flow is advected through the photosphere. Each time-bin can be thought of being the result of a separate pancake emitted by the central engine and that releases its photons at the photosphere. Each pancake has different initial parameters at the central engine, such as luminosity $L$, mass outflow rate $\dot{M}$, and initial radius, $r_0$: The dimensionless entropy of the pancake thus is given by $\eta = L/\dot{M} c^2$.
The ratio
\begin{eqnarray}
\frac{r_{\rm ph}}{r_{\rm s}} = \frac{L\sigma_{\rm T}}{ 4 \pi m_{\rm p} c^3 \eta^2 \Gamma^2 r_0} \propto \frac{L}{r_0}\,\,\eta^{-4} \hspace{5mm} r_{\rm ph} > r_{\rm s} \nonumber \\
\frac{r_{\rm ph}}{r_{\rm s}} = \left( \frac{L\sigma_{\rm T}}{12 \pi m_{\rm p} c^3 \eta^4 r_0} \right)^{1/3} \propto \left( \frac{L}{r_0} \right)^{1/3} \eta^{-4/3} \hspace{5mm} r_{\rm ph} < r_{\rm s}
\label{eq:rr}
\end{eqnarray}

\noindent \citep{Rees&Meszaros2005}.

The strong dependence on $\eta$  suggests that it is the main cause for the variation in $r_{\rm ph}/r_{\rm s}$. The observed increase of this ratio thus requires a decreasing dimensionless entropy of the flow, $\eta$.
We further argue that since $L$ does not vary much (as seen from the light curve), the change in $\eta$ can be naturally assigned mainly to an increase in mass outflow rate, $\dot{M}$. This can be due to added baryon pollution in the jet due to properties of the accretion onto the black hole, or due to interaction between the jet and the surrounding stellar material which could 
cause an increased mixing (\citet{Morsony2010, Nakar2017}; see further discussion in \S \ref{sec:lateral}.

\subsection{Broken power-law decay of the temperature}
\label{sec:BPL}

In GRB100705, we found that the temperature decays like a broken power-law. 
Interestingly, the ratio $\etaeta$ reached unity at the break-time of the $kT$ evolution, see Figure
\ref{fig:GRB100705}. This suggests that the approximately constant temperature  before the break is due to the emission occurring during the acceleration phase. 

This can be explained by the following two facts. First, for any value of $r_{\rm ph} << r_{\rm s}$, the observed temperature should indeed be constant $kT = \Gamma(r_{\rm ph}) kT'(r_{\rm ph}) = T_0$, since the comoving temperature $kT' \propto r^{-1}$ and $\Gamma \propto r$. Second, since $T_0 \propto L_0^{1/4}$, $T_0$ is only weakly dependent on the temporal variations in the fireball luminosity, $L_0(t)$. In addition, $L_0$ does not vary much, as indicated by the light-curve. In conclusion, the observed temperature is expected to be largely independent of the ratio $r_{\rm ph}/r_{\rm s}$ during the acceleration phase, even as this ratio varies for different analysed time-bins. 


The observed cooling after the break, thus occurs as the photosphere moves into the coasting phase of the jet ($\etaeta \lesssim 1$). The spectrum attains a broader shape approaching the black solid-line  spectrum in Figure \ref{fig:1}.
Further, assuming that the mass outflow rate continues to increase, the ratio $r_{\rm ph}/r_{\rm s}$ will increasingly grow (Eq. \ref{eq:rr}) for consecutive pancakes of the flow.  Since each individual pancake cools adiabatic in the coasting regime, following $T \propto (r/r_{\rm s})^{-2/3}$, the  increasing ratio $r_{\rm ph}/r_{\rm s}$ will produce the observed decrease in temperature.

Assuming a simple toy model with $\eta(t) \propto t^{s}$ for $\eta < \eta*$ in combination with the observed temperature decays of $T_{\rm obs}(t) \propto t^{b_{\rm T}}$, then Equation (\ref{eq:rr}) and adiabatic cooling yield that  $ b_{\rm T} = 8 \, s/3$. In such a case, GRB101219, with $b_{\rm T} = -0.4 \pm 0.22$ (Fig. \ref{fig:GRB100705}) would have $s \sim -0.15$, while the steeper temperature decay in GRB110721A ($b_{\rm T}-1.66 \pm 0.15$  ; \citet{Axelsson2012}) would give $s \sim -0.62$. Since $\Gamma = \eta$ in the coasting phase, such values of $s$ are consistent with observations of $\Gamma(t)$ decays in other bursts \citep[e.g.][]{Iyyani2015, Iyyani2016}).
We also note that by extending this toy model to $\eta > \eta_*$ (below $r_{\rm s}$) the temporal variation in $r_{\rm ph}/r_{\rm s}(t)$ would be slower (Eq. \ref{eq:rr}; the temporal power-law index would be smaller by a factor of 1/3). 

Indeed, the broken power-law cooling was shown by \citet{RydePeer2009} to be a recurring behaviour in bursts that have a detected thermal component. They found the pre- and post-break power-law indices to have average values of $a_{\rm T} = -0.07 \pm 0.19$ and $b_{\rm T} = -0.68 \pm 0.24$. They found significant dispersions which reflect variation between individual bursts. The dispersion in the parameter $b_{\rm T}$ was [-0.3 -- -1.3]. This dispersion can thus, according to the above, be interpreted as being caused by differences, between individual bursts, in the behaviour of temporal variation of $\eta(t)$. In the toy model above ($\eta(t) \propto t^{s}$) this corresponds to a variation in the value of $s$. The averaged value of the (post-break) temperature decay in the \citet{RydePeer2009}-sample of $<b_{\rm T}> = -0.68$ corresponds to $s = -0.25$.  

The broken power-law behaviour of the observed temperature is thus suggested to be caused by a combination of two effects. First, the transition of the photosphere from the acceleration phase to the coasting phase gives the break in the temperature evolution. Second, the flow properties (most notably $\eta = \eta(t)$) determines the slopes before and after the break. 
We point out that another interpretation of the temperature decay has been given in \citet{Peer2008, RydePeer2009}, where 2D effects of the photosphere was taken into account.




\subsection{{Limits on energy dissipation}}

The narrow observed spectra implies limits on energy dissipation within the jet. In case the photosphere occurs before the jet saturates, it is hard to dissipate much energy simply because the jet energy budget is dominated by  the photons; this implies $\delta u_\gamma \ll u_\gamma$, where $u_\gamma$ is the energy density of photons before dissipation occurs and $\delta u_\gamma$ is the amount of dissipated energy (per unit volume).

On the other hand, dissipation occurring in the coasting phase can modify the spectrum. The fate of the dissipated energy depends on if the photon spectrum has time to thermalize after the dissipation event (or rather, approach kinetic equilibrium with the electrons) before reaching the photosphere. The Compton $y$-parameter after the dissipation event is $y \approx 4 \theta \tau_{\rm d}$, where $\theta = kT/m_e c^2$ is the plasma temperature after dissipation and $\tau_d$ equals the typical number of scatterings that a photon undergoes before escaping from the dissipation optical depth $\tau_{\rm d}$. {For dissipation in the coasting phase $\tau_{\rm d} \lesssim \tau_{\rm sat} \equiv r_{\rm ph}/r_{\rm s}$.}  The $y$-parameter equals unity at $\tau_W \approx (4\theta)^{-1}$ (defining the outer boundary of the Wien zone \citet{Beloborodov2013}). Significant dissipation can occur at $\tau_d \gg \tau_W$ without broadening the spectrum emitted at the photosphere, as the photon spectrum will simply thermalize into a Wien spectrum at the new temperature before being emitted. However, any dissipation occurring further out in the jet {($\tau_d \lesssim  \rm{min} [ \tau_W, \tau_{\rm sat}]$)} is limited in amount to $\delta u_\gamma \ll u_\gamma$ in order to not broaden the observed spectrum. This is quite constraining far out in the coasting jet phase, where we have $u_\gamma / \rho c^2 \approx (r/r_s)^{-2/3} \ll 1$ for $r \gg r_s$, and where $\rho$ is the mass density. This implies that dissipating a fraction $\sim (r/r_s)^{-2/3} \ll 1$ of the kinetic energy is enough to have $\delta u_\gamma \sim u_\gamma$ and modify the spectrum. Any spectra with $\eta/\eta_\star \ll 1$, which are emitted far in the coasting phase, must be emitted by very ``smooth'' jets in order to not have their spectra broadened by dissipation.

\subsection{{Limits on the jet variability time}}

The time-bin size $\Delta t$ used for the spectral fitting is of order a few seconds. One could expect the GRB spectrum to change on a much smaller time $\delta t$, broadening the observed spectrum (which is integrated over time $\Delta t$). The fact that the spectra presented in this work are not broadened significantly during this time interval puts constraints on the jet variability (for these specific GRBs).

A natural time scale for variations is the light crossing time of the central engine, $R/c \sim 3 \times 10^{-4}$~s, where $R \sim 10^7$~cm is the central engine size. Time bins of $\Delta t \sim 3$~s correspond to $\sim 10^4$ light crossing times, implying a very steady central engine. Any central engine variability on times $\delta t \ll \Delta t$ must be small enough to not dissipate significant energy outside the Wien zone.

The fits to GRB100507 indicate that the jet base is located at $r_0 \sim 3 \times 10^9$~cm (see Eq. \ref{eq:ro}). This could be consistent with free jet expansion beginning at $r_0 \lesssim R_{star}$, where $R_{star}$ is the radius of the stellar (Wolf-Rayet) progenitor; at smaller radii the jet would propagate at mildly relativistic speeds, being collimated by collimation shocks from the surrounding cocoon. The associated time scale is $\delta t \sim r_0/c \sim 0.1$~s, possibly washing out any central engine variability on shorter time scales. {However, numerical simulations of the jet propagation through the star by \citet{Morsony2010} indicate that any short time-scale variations introduced by the central engine actually survive unmodified within the jet as it emerges from the star.} In any case, the jet base must have been steady on time scales  $(\Delta t/\delta t) \delta t \sim 30 \, r_0/c$.

\subsection{Radiative efficiency and structure of the jet}
\label{sec:radeff}



As mentioned in \S \ref{sec:GRB101219}, for GRB101219 there is 
{an apparent} contradiction between highly efficient prompt phase and the inferred radiative efficiency using the estimate of the blast wave energy from the afterglow. A way to reconcile these observations  is to have  a structured jet with an energy density either varying with radius (\ref{sec:evolving}) or with angle (\ref{sec:lateral}), or an additional jet component in form of a inert Poynting flux (\ref{sec:poynting}). 

\subsubsection{Evolving jet} 
\label{sec:evolving}

As suggested in \S \ref{sec:BPL} the prompt phase spectral evolution can be explained by the mass outflow rate increasing, or equivalently the dimensionless entropy $\eta$ decreasing (see Eq. \ref{eq:rr}). A decrease in $\eta$  leads to that the late-time pancakes suffer greater and greater adiabatic losses: Since $\rph/\rs$ becomes large, the thermal emission from the photosphere will become increasingly weaker and cooler, whilst most of the burst energy will be in the form of kinetic energy of the blast wave.  Without any internal dissipation that can tap the kinetic energy of the flow and transform it into radiation, the GRB will thus eventually be able to become dark in $\gamma-$rays. At the same time, the kinetic energy of the blast wave will be responsible for the emission during the afterglow. This scenario thus assumes that the central engine has a continuous activity, for instance by a slowly varying luminosity, $L_0$ even beyond the $\gamma$-ray emission has ended. Such a scenario can therefore resolve the radiative efficiency problem in GRB101219: The kinetic energy measured in the afterglow is  from parts of the jet with low $\eta$ which deposits all its energy in kinetic energy of the blast wave and has no detectable gamma-ray emission. The prompt emission is from an earlier ejected part of the jet which has a large $\eta$ and therefore releases most of its energy during the prompt phase as photospheric emission and has very little energy left to sustain an afterglow emission. 

In such a scenario, a consequence of the decreasing $\eta$, would be a formation of a plateau in the afterglow light-curve. As the photospheres of the late-time pancakes move into the coasting phase, the decreasing $\eta$ will translate into a decreasing $\Gamma$ of the blastwave. 
As the leading pancakes with large Lorentz factors start to decelerate, the trailing pancakes, having lower $\Gamma$, will catch up and  cause an energy injection into the  blast wave, thereby creating the plateau. The temporal slope will reflect the change in $\Gamma$ \citep{Sari2000, Laskar2015}.

Interestingly, GRB101219 has such a plateau, which makes the afterglow consistent with a central engine producing a flow with a decreasing Lorentz factor, which was given by the prompt spectral evolution. The plateau phase starts at 4100s and lasts until 58000s. The plateau has a temporal slope of $\propto t^{-0.46}$, which changes into a slope $\propto t^{-0.74}$ \citep{Larsson2015}. Moreover, the spectral slope of the X-ray data has a photon index\footnote{Determined by using tools supplied by the UK Swift Science Data Centre} of $\sim 2.0$. Using the standard closure relations \citep{Zhang2006} we find that the X-rays are above the cooling and minimum injection frequencies $(\nu_{\rm X} > \nu_{\rm c} > \nu_{\rm m})$. Employing the methods in \citet{Laskar2015}, the temporal slope of the plateau light curve translates into $\Gamma \propto t^{-0.24}$. This should be compared to $\eta = \Gamma \propto t^{-0.15}$, which was found from spectral modelling of the prompt phase. These two evolutions are remarkably similar. The suggestive connection between inferred decay of Lorentz factor from the prompt phase and the plateau flux slope should be further explored in other bursts.




{\subsubsection{Lateral jet} 
\label{sec:lateral}

It was early realised that as the jet drills thought the star a hot cocoon of shocked jet material is formed surrounding the jet (Ramirez-Ruiz et al. 2002; Lazzati \& Begelman 2005).  A two component jet structure is thus formed, with a narrow jet and a broader shocked jet cocoon surrounding it. Due to the viewing angle of the observer, either the jet or the shocked jet cocoon emission is observed \citep{Morsony2007,Lopez-Camara2013, Nakar2017,DeColle2017}. 

A possibility is thus that the thermal components, observed in the bursts analysed above, come from the shocked jet cocoon.  The prompt emission from the  jet itself is not detected in the GBM energy range, since the angle between the outer edge of the (unshocked) jet and the line-of-sight is larger than $\Gamma^{-1}$, i.e. the burst is viewed at a large viewing angle. At the same time, the jet kinetic energy would be revealed from the afterglow observations, since the jet afterglow is detectable from larger viewing angles compared the prompt jet emission, due to the deceleration of the jet. The kinetic energy estimated from the  afterglow observations is thus, to a large extent, related to the jet kinetic energy and not directly connected to observed prompt emission, which is from the cocoon. The large viewing angle, suggested in this scenario, is indeed consistent with the fact that a clear jet break in the afterglow emission is not detected \citep{Larsson2015}, since for an off-axis viewer the jet-break is expected to be smeared out and the full transition to be delayed \citep{vanEerten2010,vanEerten2011}. 

As the jet propagates through the star, internal energy is deposited in the shocked jet cocoon. A mixing of stellar material into the shocked jet cocoon could take place thereby changing the baryon load of the shocked jet cocoon.
The observations of the bursts in this paper require, however, that the photosphere occurs below the saturation radius. Therefore, the mixing must be small to ensures that the $\eta$-value of the jet cocoon is (initially) high: Without any mixing $\eta$ of the shocked jet cocoon would be similar to that of the jet \citep{Nakar2017}.   
Such a shocked jet cocoon behaves similarly to the jet itself. The main differences are that the cocoon is expected to have a large opening angle, assumed to be $\theta_c \sim 0.5$,  and its size of the launch base, or similarly $r_0 \sim r_* \theta_j$, where the $\theta_j \sim 0.2$ is the canonical opening angle of the jet as it emerges from the star. 
Moreover, numerical simulations seem to suggest that  the material of the shocked jet cocoon  that is closest to the head of the jet has experiences less mixing compared to material further into the star. An increasing mixing could thus explain the decreasing value of $\eta$ imposed by the observed bursts. 
Therefore, similarly to the assumed variations of the flow due to central engine variability in \S\ref{sec:evolving}, the variation in  mixing throughout the shocked jet cocoon could be the cause of the variations in observed spectral properties. 

The shocked jet cocoon is also expected to give rise to an afterglow component, but it is much weaker than the afterglow generated by the jet itself \citep{Nakar2017}.

The duration of the observed thermal component is in this scenario connected to the duration of the cocoon emission, which  is limited by the scale of the progenitor star, $r_*$. For the limiting case of no mixing at all $t\sim r_*/c = r_0/ \theta_j/c \sim 1 $s, with $r_0 = 5 \times 10^9$ (estimated for the analysed bursts) which is an orders of magnitude shorter than the observed durations. However, partial mixing and an evolving $\eta$ could explain the longer durations, but this needs to be investigated by numerical modelling.

Moreover, the scenario also raises the question of why so few bursts are observed to have a single thermal component. Since the cocoon is expected to subtend a large angle, many such bursts should have been detected. A possibility is that a large mixing typically occurs between the stellar and jet material. The cocoon emission is then expected to be soft and weak and thereby not detected. The bursts analysed in this paper are thus non-typical in that they exhibit very low mixing. 
 

Finally, we note that the blackbody component in the early X-ray afterglow of GRB101219 was interpreted as emission from the cocoon (\citet{Starling2012}, but in a scenario with high baryon load, low $\eta$ and internal shocks \citep{Peer2006}. 

\subsubsection{Magnetic content of the jet}
\label{sec:poynting}

{A third possibility is that the jet includes a significant Poynting flux component at the base of the flow in addition to a thermal component, the so called hybrid model \citep[e.g.][]{Zhang&Meszaros2002, ZhangPeer2009, Guiriec2011, Hascoet2013, Iyyani2013}.  If the Poynting flux does not contribute to the prompt emission, the only emission will be that from the photosphere, which will simply emit the thermal component of the jet. The Poynting flux is later available to power the afterglow emission (Lyutikov \& Blandford 2003; Mimica et al. 2009). 

In hybrid flows the thermal acceleration ($\Gamma \propto r^{1}$) is assumed to proceed any magnetic acceleration \citep{VK2003}. The latter could, for instance, be caused by magnetic reconnection in the flow, and occurs at a slower rate, 
typically assumed to follow $\Gamma \propto r^{1/3}$ \citep{Drenkhahn&Spruit2002, Begue2017}. 
 
The observations of the bursts presented in this paper put two requirements on such a scenario. 

(i) The narrowness of the spectra requires $\Gamma \propto r^{-1}$. 
This means that the photosphere has to occur below the thermal saturation radius and before any magnetically driven acceleration dominates. Moreover, if the photosphere were to occur during the $\Gamma \propto r^{1/3}$ phase, a high energy component due to Comptonization would appear above the peak \citep{Giannios2006}, and the peak would eventually be shifted to very large energies $> 8$ MeV due to photon starvation \citep{BeguePeer2015}. These facts also point to the observed photosphere occurring during the thermal acceleration phase.

(ii) Since the magnetic component does not contribute to the prompt emission, the amount of dissipation that occurs must be low, either by magnetic reconnection or internal shocks.
The absence of significant magnetic reconnection limits the acceleration that the magnetic field can perform on the flow. Such an inert magnetic field will simply be advected with the flow and be deposited in the afterglow.  Moreover, internal shocks are indeed expected to be weak in highly magnetized flows (Zhang \& Kobayashi 2005) which thus is a natural reason for the paucity of prompt emission from the magnetic component.


Finally, we note that the hybrid model has been suggested for bursts in which both a thermal and a non-thermal component have been observed \citep[e.g.][]{Ryde2005, Battelino2007, RydePeer2009, Guiriec2011, Iyyani2013, Axelsson2012, Burgess2014a, Nappo2017}. In these cases the magnetic energy is dissipated during the prompt phase, giving rise to emission observed in coincidence with the photospheric emission \citep[e.g.][]{Meszaros&Rees2000, Daigne&Mochkovitch2002}. The efficiency of such emission can, for instance, be solved by invoking the ICMART (Internal-Collision- Induced Magnetic Reconnection and Turbulence) process (Zhang \& Yan 2011).


}





%

%


\section{Conclusion}

We have analysed GRB100507 and GRB101219 which have very narrow spectra. We have shown that such spectra imply a high radiative efficiency: the photospheres occur in the jet acceleration phase. They also imply that the jet energy must be dominated by radiation as opposed to magnetic fields. In addition, such spectra set constraints on the dissipation profile - very little dissipation is allowed outside the Wien zone. They also set constraints on the time variability of the jets, since the spectra are not smeared out during the time-bins.

Based on the spectral evolution in GRB100507, we have argued that the mass entrainment increases with time. This implies that the photosphere will reach and pass the saturation radius, causing two temporal phases: initially, the spectra are narrow and have a (close to) constant temperature. Later, the spectra become broader and the temperature starts to decrease. 

Finally, we have argued that the observations of GRB101219 can be explained by  an undisruptive jet which becomes dark in $\gamma$-rays, while producing an energetic blast wave, leading to an afterglow with a plateau phase. Alternatively, the thermal $\gamma$-ray emission could be from a cocoon surrounding the jet. If the latter scenario is the correct interpretation, the very narrow spectra observed, indicate the existence of a shocked jet cocoon, which initially has a very low mixing of stellar material.



\section*{Acknowledgements}

We thank Drs. Damien B\'egu\'e, J. Michael Burgess, Josefin Larsson, and Liang Li for enlightening discussions. This research has made use of data obtained through the High Energy Astrophysics Science Archive Research Center Online Service, provided by the NASA/Goddard Space Flight Center. We acknowledge support from the Swedish National Space Board and the Swedish Research Council (Vetenskapsr{\aa}det). FR is supported by the  G\"oran Gustafsson Foundation for Research in Natural Sciences and Medicine. 



\bibliographystyle{mnras}
\bibliography{ref2017} 




\appendix

\section{The narrowest GRB spectrum (equation~2)}

All lab frame quantities (except for the radius) are here denoted by a tilde sign. The spectral flux which propagates outward through the lab frame sphere of radius $r$ is $\Fl = \int \Il \mul \D\Omegal$, where $\mul$ is the cosine of the angle to the local radial direction, and the integration is over the outer half-sphere, i.e. radiation which propagates outwards, $0 \leq \mul \leq 1$. The lab frame spectral luminosity is $\Ll = 4\pi r^2 \Fl$, or

\be
\Ll = 4\pi r^2 \int \Il\mul\D\Omegal.
\ee

\noindent We aim to find $\nul \Ll / \Lintl$, where $\Lintl \equiv \int \Ll \D\nul$. This will give a normalized, dimensionless lab frame $\nu F_\nu$ spectrum (see also \citet{begue2013monte}).

\citet{Beloborodov2011} showed that the lab frame intensity does not change with radius as long as the jet is in the acceleration phase ($\Gamma \propto r$), and so if the photons decouple during the acceleration phase, this will also be true for the observed spectrum for an observer located at infinity.


Scatterings, emission and absorption ensures that the comoving intensity in the deep accelerating phase is blackbody (and isotropic), such that $\I = C \nu^3/[\exp(h\nu/kT) - 1]$, where $C$ is simply a constant (which depends on the temperature, or energy density of the plasma). The lab frame frequency is related to the comoving frame frequency by the Doppler boost, $\nul = D\nu$, and the intensity transforms as $\Il = D^3\I$, and so

\be
\Ll = (4\pi)^2 r^2 C \frac{1}{2} \int\limits_0^1 \frac{\nul^3 \mul}{\exp\left(\frac{h\nul}{DkT}\right) - 1} \D\mul.
\label{eq:2}
\ee

\noindent $\Lintl$ is then obtained by first integrating over the lab frame frequency, and then performing the angular integral. The frequency integration is performed by changing variable to $y \equiv h\nul/DkT$, so that $\int \nul^3/[\exp(h\nul/DkT)-1] \D\nul = (DkT/h)^4 \int y^3/[\exp(y)-1] \D y = (\pi^4/15) (DkT/h)^4$. We then have

\be
\Lintl = \frac{\pi^4}{30} (4\pi)^2 r^2 C \left(\frac{kT}{h}\right)^4 \int\limits_0^1 D^4 \mul\D\mul.
\ee

\noindent We know that $\mul = (\mu + \beta)/(1 + \beta\mu)$, and $\D\mul = D^{-2}\D\mu$. The limit $\mul = 0$ then becomes $\mu = -\beta$. We find

\be
\int\limits_0^1 D^4 \mul\D\mul = \int\limits_{-\beta}^1 D^2 \frac{\mu + \beta}{1 + \beta\mu} \D\mu.
\ee

\noindent Now, using the fact that $D = \Gamma(1+\beta\mu)$, and then taking the formal limit $\beta \rightarrow 1$, we find that the above integral equals $(8/3)\Gamma^2$, and

\be
\Lintl = \frac{4}{3} \frac{\pi^4}{15} \frac{(4\pi)^2 r^2 C}{\Gamma^2} \left(\frac{\Gamma kT}{h}\right)^4.
\ee

Now, going back to equation (\ref{eq:2}), we change the variable of integration from $\mul$ to $\mu$, and then from $\mu$ to $u \equiv 1+\beta\mu$, take the limit $\beta \rightarrow 1$, and define a new energy variable $x \equiv h\nul/\Gamma kT$ (energy normalized to the ``boosted temperature''). We then have

\be
\frac{\nul\Ll}{\Lintl} = \frac{45}{8\pi^4} x^4 \int\limits_0^2 \frac{u^{-2}}{\exp(\frac{x}{u}) - 1} \D u.
\label{eq:spectrum1}
\ee

\noindent This expression is similar to a blackbody (which would be proportional to $x^4/(\exp(x)-1)$, but it instead has $\exp(x/u)$, where $u$ varies, and so the spectrum consists of ``many temperatures''. Further changing the variable to $w \equiv \exp(x/u) - 1$, we find

\be
\int\limits_0^2 \frac{u^{-2}}{\exp(x/u)-1} \D u = \frac{1}{x} \int\limits_{\exp(x/2) - 1}^\infty \frac{1}{w(w+1)} \D w =\newline 
\frac{1}{x}\left\{\frac{x}{2} - \ln\left[\exp\left(\frac{x}{2}\right) - 1\right]\right\},
\ee

\noindent so that finally, we arrive at

\be
\frac{\nul\Ll}{\Lintl} = \frac{45}{8\pi^4} x^3 \left\{\frac{x}{2} - \ln\left[\exp\left(\frac{x}{2}\right) - 1\right]\right\},
\label{eq:spectrum2}
\ee

\noindent which is eq. (\ref{eq:aBB}) and is the narrowest spectrum that can be observed from a GRB. 





\bsp	
\label{lastpage}
\end{document}